\documentclass[english]{article}
\usepackage[T1]{fontenc}
\usepackage[latin9]{inputenc}
\usepackage{geometry}
\usepackage{units}
\usepackage{amsmath}

\makeatletter
\numberwithin{equation}{section}

\makeatother

\usepackage{babel}
\begin{document}
\title{An Optimal Weighting Function for the Savitzky-Golay Filter}
\author{Paul W. Oxby}
\date{2021 November 22}
\maketitle
\begin{abstract}
The Savitzky-Golay FIR digital filter is based on a least-squares
polynomial fit to a hypothetical sample of equally spaced data. This
gives the filter the ability to preserve moments of features like
peaks in the input. Descriptions of the filter typically consider
the case where equal weights are implicitly applied to the residuals
of the fit. In a largely overlooked paper Turton showed that weighting
the residuals with a triangular function significantly improves the
frequency response of the filter in the stopband. The Savitzky-Golay
filter is commonly referred to as a smoothing filter. This paper uses
a particular measure of smoothness to show that a quadratic residual
weighting function optimizes the smoothness of the filter output for
a given sample size and degree of the fitting polynomial. This weighting
function can provide substantially better smoothness than that with
a constant weighting function.
\end{abstract}

\section{Introduction}

In a 1964 paper Savitzky and Golay {[}1{]} showed how a least-squares
polynomial fit to a sample of equally spaced data could be used as
the basis of an FIR digital filter to smooth the noisy data typically
generated by chemical analysis instruments like spectrometers. This
paper has the distinction of being one of the most cited papers in
the field of analytical chemistry. But as Schafer {[}2{]} has observed,
the Savitzky-Golay (S-G) filter is not widely known within the digital
signal processing community. This is partly because the frequency
response of the S-G filter in the stopband region is mediocre. In
a 1992 paper Turton {[}3{]} showed that weighting the least-squares
residuals with a triangular function significantly improves the frequency
response of the S-G filter in the stopband. However citation indices
give relatively few citations for his paper which suggests that it
merits more attention than it has received. This paper extends Turton's
analysis by showing that a quadratic weighting function optimizes
a particular time-domain measure of the smoothness of the output of
the S-G filter.

$ $

The material in this paper involves two disciplines, statistics and
signal processing. The statistics term ``measurement error'' roughly
corresponds to the signal processing term ``noise''. For the purpose
of this discussion it will be assumed that the measurement error or
noise is an independent and identically distributed (i.i.d.) random
variable with zero mean and that the true measurements or signal are
smooth enough to be locally represented by a polynomial. Where this
paper discusses statistical concepts and theory the terms ``measurement''
and ``error'' are used.\newpage{}

\section{The Savitzky-Golay Smoothing Filter}

This section shows how the Savitzky-Golay (S-G) filter is derived
from the statistical theory of the optimal polynomial fit to a sample
of measurements corrupted by random measurement error. It is assumed
that \emph{q} values of a hypothetical dependent variable, \emph{y},
are associated with \emph{q} equally spaced values of an independent
variable, \emph{x,} and that the relationship between \emph{x} and
\emph{y} can be represented by a polynomial over the range of the
\emph{q} samples. To illustrate, if a quadratic equation in \emph{x}
with three unknown parameters, $a_{1}$, $a_{2}$, and $a_{3}$ is
to be fit to five corresponding measurements of \emph{y} then this
can be expressed by the following set of five equations in three unknowns:
\begin{align}
y_{1} & =a_{1}+a_{2}x_{1}+a_{3}x_{1}^{2}\nonumber \\
y_{2} & =a_{1}+a_{2}x_{2}+a_{3}x_{2}^{2}\nonumber \\
y_{3} & =a_{1}+a_{2}x_{3}+a_{3}x_{3}^{2}\\
y_{4} & =a_{1}+a_{2}x_{4}+a_{3}x_{4}^{2}\nonumber \\
y_{5} & =a_{1}+a_{2}x_{5}+a_{3}x_{5}^{2}\nonumber 
\end{align}
The number of unknown parameters, \emph{a}, will be denoted by \emph{n}.
So in this case $q=5$ and $n=3$. Using matrix notation Equations
2.1 are more compactly expressed as:

\begin{equation}
Xa=y
\end{equation}
where \emph{X} is a $5\times3$ matrix the first column of which is
a vector of ones, the second column is the vector \emph{x} and the
third column is the vector of the squares of \emph{x}. For the purpose
of deriving a smoothing filter the scale of \emph{x} doesn't matter
but it will be assumed that the values of \emph{x} are equally spaced.

Because there are more equations for the elements of vector \emph{a}
than there are elements of \emph{a} there won't be a solution for
\emph{a} that exactly satisfies all of the equations if there is random
error associated with the dependent variable, \emph{y}. The system
of equations is said to be over-determined. The classic least-squares
solution to a system of overdetermined linear equations gives values
of the elements of the vector \emph{a} that minimizes the sum of squares
of the elements of the residual vector, $\rho$, given by:

\begin{equation}
\rho=Xa-y
\end{equation}

This optimal values of the elements of vector \emph{a} are obtained
by premultiplying both sides of Equation 2.2 by the transpose, $X^{T}$,
of the matrix $X$ which yields a matrix equation whose solution minimizes
the sum of squares of the residuals which can be expressed as $\rho^{T}\rho$:

\begin{equation}
a=\left(X^{T}X\right)^{-1}X^{T}y
\end{equation}
The values of the polynomial fit are given by the vector $\widehat{y}$
which is the best estimate of the true values of the measurements:

\begin{equation}
\widehat{y}=Xa=X\left(X^{T}X\right)^{-1}X^{T}y
\end{equation}
If a particular element of the $\widehat{y}$ vector is of interest,
say the \emph{j}th element, then that element can be isolated from
$\widehat{y}$ by a simple matrix operation. The vector \emph{u} is
constructed with \emph{q} elements all of which are set to zero except
for the \emph{j}th element which is set to one:

\begin{equation}
u_{i}=\left\{ \begin{array}{l}
0\;\mathrm{for}\:i\neq j\\
1\;\mathrm{for}\:i=j
\end{array}\right.\quad(\mathrm{for}\:i=1\,...\,q)
\end{equation}

The \emph{j}th element of $\widehat{y}$ is now given by:

\begin{equation}
\widehat{y}_{j}=u^{T}\widehat{y}=u^{T}X\left(X^{T}X\right)^{-1}X^{T}y
\end{equation}
\newpage The transpose of $u^{T}X\left(X^{T}X\right)^{-1}X^{T}$ is
a vector of \emph{q} filter coefficients, \emph{c}, which is independent
of the elements of the vector \emph{y}:

\begin{equation}
c=X\left(X^{T}X\right)^{-1}X^{T}u
\end{equation}

Therefore the filtering operation is simply the dot product or convolution
of the filter coefficients, \emph{c}, with the measurements, \emph{y:}

\begin{equation}
\widehat{y}_{j}=c^{T}y
\end{equation}
The last two equations define the standard S-G smoothing filter. If
the filter parameters, \emph{q} and \emph{n}, are selected to avoid
overfitting or underfitting the values of the filter input then features
of the filter input that can be represented by the fitting polynomial
will pass through the filter without distortion.

$ $

To summarize, the elements of the vector $\widehat{y}$, given by
the polynomial fit, Equation 2.5, are the best estimates of the true
values of the elements of the measurement vector \emph{y}. The \emph{j}th
element of the vector \emph{u} in Equation 2.7 is set to one therefore
$u^{T}\widehat{y}$ selects the \emph{j}th element of $\widehat{y}$,
the scalar $\widehat{y}_{j}$. Statistical estimation theory shows
that the optimal value of \emph{j} corresponds to the middle element
of $\widehat{y}$ with index $\nicefrac{(1+q)}{2}$ and for which
$u_{(1+q)/2}=1$. When $j=\nicefrac{(1+q)}{2}$ then \emph{c} is a
symmetric vector and the filter is linear phase. In this case the
value of \emph{q} must be odd. 

$ $

This standard description of the S-G filter implicitly assigns equal
weights to the residuals, $\rho$, in calculating the sum of squares
of the residuals which can be expressed as $\rho^{T}\rho$. A weighted
sum of squares of the residuals can be expressed as $\rho^{T}W\rho$
where \emph{W} is a $q\times q$ matrix of weights. In statistical
estimation theory the optimal weight matrix \emph{W} is the inverse
of the error variance-covariance matrix. Because the measurement errors
are assumed to be independent the weight matrix \emph{W} can be taken
to be diagonal. The value of the filter coefficient vector, \emph{c},
that minimizes the weighted sum of squares of the residuals, $\rho^{T}W\rho$,
is given by:
\begin{equation}
c=WX\left(X^{T}WX\right)^{-1}X^{T}u
\end{equation}

The derivative of the vector \emph{c} with respect to the diagonal
element $W_{k,k}$ is given by:

\begin{equation}
\dfrac{dc}{dW_{k,k}}=\left[I-WX\left(X^{T}WX\right)^{-1}X^{T}\right]\dfrac{dW}{dW_{k,k}}X\left(X^{T}WX\right)^{-1}X^{T}u
\end{equation}

where \emph{I} is a $q\times q$ identity matrix and \emph{$\nicefrac{dW}{dW_{k,k}}$}
is a $q\times q$ diagonal matrix with only one nonzero element which
is one:

\begin{equation}
\left[\dfrac{dW}{dW_{k,k}}\right]_{i,i}=\left\{ \begin{array}{l}
0\;\mathrm{for}\:i\neq k\\
1\;\mathrm{for}\:i=k
\end{array}\right.\quad(\mathrm{for}\:i=1\,...\,q)
\end{equation}
\newpage{}

\section{The Smoothness of the Output of an FIR Filter}

The last section described the standard (i.e., equal residual weights)
and the weighted version of the S-G smoothing filter. The smoothness
of the output of a general FIR filter will now be quantified. If the
impulse to the filter represents a unit input of measurement error
(noise) then the filter distributes the unit impulse of error over
an output response vector equal to the coefficient vector. This distribution
of an unit input impulse of error over an output vector response can
be described as a smoothing operation.

$ $

The output of a general FIR filter, $\mathbf{y}$, is a convolution
of the filter coefficients and the filter input:

\begin{equation}
\mathbf{y}=c_{1}y_{1}+c_{2}y_{2}+\cdots+c_{q}y_{q}
\end{equation}

The values of the filter input, $y_{i}$, can be considered as having
two additive components, the ``true'' values of the input and the
errors. Because the filtering transformation is linear the filter's
action on the input errors can be analysed independently of the ``true''
values of the input. The error component of the filter input will
be denoted by \emph{e }and the error component of the filter output,
$\mathbf{e}$, is given by:

\begin{equation}
\mathbf{e}=c_{1}e_{1}+c_{2}e_{2}+\cdots+c_{q}e_{q}
\end{equation}

It is assumed that the values of the input error, $e_{i}$, are uncorrelated
random variables with a mean of zero and constant variance, $\sigma_{e}^{2}$.
With this assumption the variance of the output error, $\mathbf{e}$
, is given by:

\begin{equation}
\sigma_{\mathbf{e}}^{2}=c_{1}^{2}\sigma_{e}^{2}+c_{2}^{2}\sigma_{e}^{2}+\cdots+c_{q}^{2}\sigma_{e}^{2}
\end{equation}

The ratio of the variance of the filter output error to the variance
of the input error, the error reduction ratio (aka noise reduction
ratio), will be denoted by\emph{ r} and is given by:

\begin{equation}
r=\dfrac{\sigma_{\mathbf{e}}^{2}}{\sigma_{e}^{2}}=c_{1}^{2}+c_{2}^{2}+\cdots+c_{q}^{2}=c^{T}c
\end{equation}

The output error, $\mathbf{e}$, can be considered to be smooth to
the extent that the difference between successive values of the output
error is small. The difference between successive values of the output
error is:

\begin{equation}
\Delta\mathbf{e}=\left(c_{1}e_{2}+c_{2}e_{3}+\cdots+c_{q}e_{q+1}\right)-\left(c_{1}e_{1}+c_{2}e_{2}+\cdots+c_{q}e_{q}\right)
\end{equation}

Rearranging terms and padding with zeroes gives:

\begin{equation}
\Delta\mathbf{e}=\left(0-c_{1}\right)e_{1}+\left(c_{1}-c_{2}\right)e_{2}+\cdots+\left(c_{q-1}-c_{q}\right)e_{q}+\left(c_{q}-0\right)e_{q+1}
\end{equation}

Padding with zeroes makes this equation symmetric with respect to
all of the filter coefficients, particularly $c_{1}$ and $c_{q}$.
The variance of the difference between successive values of the output
error is:
\begin{equation}
\sigma_{\Delta\mathbf{e}}^{2}=\left(0-c_{1}\right)^{2}\sigma_{e}^{2}+\left(c_{1}-c_{2}\right)^{2}\sigma_{e}^{2}+\cdots+\left(c_{q-1}-c_{q}\right)^{2}\sigma_{e}^{2}+\left(c_{q}-0\right)^{2}\sigma_{e}^{2}
\end{equation}

The ratio of the variance in the difference between successive values
of the output error to the variance in the difference between successive
values of the input error will be denoted by \emph{s} given by:

\begin{equation}
s=\dfrac{\sigma_{\Delta\mathbf{e}}^{2}}{\sigma_{\Delta e}^{2}}=\dfrac{1}{2}\left[\left(0-c_{1}\right)^{2}+\left(c_{1}-c_{2}\right)^{2}+\cdots+\left(c_{q-1}-c_{q}\right)^{2}+\left(c_{q}-0\right)^{2}\right]
\end{equation}

This ratio, \emph{s}, is a measure of the filter effectiveness in
smoothing error (noise) and will be referred to here as the filter
smoothing parameter. The factor of two is due to the fact that the
variance in the difference between successive values of the input
error is twice the variance of the input error.\newpage{} Like the
error reduction ratio of Equation 3.4, the smoothing parameter characterizes
the filter itself and is independent of the filter input. In the degenerate
case where $q=1$ and $c_{1}=1$ the filter output is equal to the
filter input and the values of both the error reduction ratio, \emph{r},
and the smoothing parameter, \emph{s}, are one. If \emph{c} were a
stochastic variable then \emph{s} would be one minus the autocorrelation
in \emph{c} or, equivalently, one minus the autocorrelation in the
filter impulse response.

The padding with zeroes in Equation 3.8 serves to emphasize an important
point. The output of an FIR filter will be smooth to the extent that
the terms $(c_{i+1}-c_{i})^{2}$ in this equation are small. When
the residuals of the S-G filter are given equal weights then the terms
$(0-c_{1})^{2}$ and $(c_{q}-0)^{2}$ tend to dominate the other terms
in value. It will be shown that an alternative residual weighting
function will reduce the influence of these two terms on the filter
smoothness.

$ $

Equation 3.8 can be expressed in compact matrix notation as:

\begin{equation}
s=\dfrac{c^{T}Tc}{2}
\end{equation}

where \emph{T} is a tridiagonal matrix whose elements are two on the
diagonal and minus one on the off-diagonals:

\begin{equation}
\begin{array}{r}
T_{i,i}=2\quad\left(\mathrm{for}\:i=1\,...\,q\right)\quad\;\;\,\\
\\
T_{i,i+1}=T_{i+1,i}=-1\quad\left(\mathrm{for}\:i=1\,...\,q-1\right)
\end{array}
\end{equation}

The derivative of the scalar \emph{s} with respect to the elements
of vector \emph{c} is given by:

\begin{equation}
\dfrac{ds}{dc}=Tc
\end{equation}

\section{The Optimal Weight Matrix For the Smoothing Parameter}

The vector \emph{v} is constructed with \emph{q} elements all of which
are one:

\begin{equation}
v_{i}=1\quad\left(\mathrm{for}\:i=1\,...\,q\right)
\end{equation}

The weight vector \emph{w} is defined as:

\begin{equation}
w=T^{-1}v
\end{equation}

where \emph{T} is given by Equation 3.10. The residual weight matrix,
\emph{W}, is diagonal and the elements of the diagonal correspond
to the elements of the weight vector \emph{w}:

\begin{equation}
W_{i,i}=w_{i}\quad\left(\mathrm{for}\:i=1\,...\,q\right)
\end{equation}

It will now be shown that the smoothing parameter\emph{ s} is minimized
with respect to the elements of the weight vector \emph{w} defined
by Equation 4.2. From Equations 4.1 and 4.3 it follows that:

\begin{equation}
Wv=w
\end{equation}

Premultiplying Equations 4.2 and 4.4 by \emph{T} gives:

\begin{equation}
TWv=v
\end{equation}

Therefore \emph{v} is an eigenvector of \emph{TW} where the corresponding
eigenvalue, $\lambda$, is one. The matrix \emph{TW} has \emph{q}
eigenvalues given by:

\begin{equation}
\lambda_{i}=\dfrac{i(i+1)}{2}\quad\left(\mathrm{for}\:i=1\,...\,q\right)
\end{equation}
\newpage{}

If the first \emph{n} eigenvectors are assembled columnwise into the
$q\times n$ matrix \emph{V} then:

\begin{equation}
TWV=V\varLambda
\end{equation}

where $\mathrm{\varLambda}$ is a diagonal matrix of the first \emph{n}
eigenvalues:

\begin{equation}
\varLambda_{j,j}=\lambda_{j}\quad\left(\mathrm{for}\:j=1\,...\,n\right)
\end{equation}

Because the matrix \emph{TW} is not symmetric the eigenvectors will
not be orthogonal. However the eigenvectors are orthogonal with respect
to the weight matrix \emph{W}. In other words the matrix $V^{T}WV$
is diagonal. If the eigenvectors are scaled to be orthonormal with
respect to \emph{W} then the scaled eigenvectors, denoted by \emph{A},
are given by :

\begin{equation}
A_{i,j}=\dfrac{V_{i,j}}{\sqrt{\left(V^{T}WV\right)_{j,j}}}\quad\left(\mathrm{for}\:i=1\,...\,q,\:j=1\,...\,n\right)
\end{equation}

Therefore:

\begin{equation}
A^{T}WA=I
\end{equation}

and

\begin{equation}
TWA=A\varLambda
\end{equation}

Now, the columns of the matrix \emph{X} in Equation 2.2 are a set
of basis vectors based on increasing powers of the vector \emph{x}.
With the eigenvalues given in the order of Equation 4.6 the first
column of the matrix \emph{A} is a constant, the second column is
a linear function of \emph{x}, the third is quadratic, etc. Therefore
the columns of the matrix \emph{A} are an alternative set of basis
vectors that have the property of being orthonormal. This means that
wherever the matrix \emph{X} appears in the equations of Section 2
it can be replaced by the matrix \emph{A}. With this substitution
Equations 2.10 and 2.11 simplify to:

\begin{equation}
c=WAA^{T}u
\end{equation}

and

\begin{equation}
\dfrac{dc}{dW_{k,k}}=\left[I-WAA^{T}\right]\dfrac{dW}{dW_{k,k}}AA^{T}u\quad\left(\mathrm{for}\:k=1\,...\,q\right)
\end{equation}

The derivative of the smoothing parameter \emph{s} with respect to
the element of the diagonal weight matrix $W_{k,k}$ is obtained by
combining Equations 3.11, 4.12 and 4.13:

\begin{equation}
\dfrac{ds}{dW_{k,k}}=\left[\dfrac{dc}{dW_{k,k}}\right]^{T}\dfrac{ds}{dc}=u^{T}AA^{T}\dfrac{dW}{dW_{k,k}}\left[I-AA^{T}W\right]TWAA^{T}u
\end{equation}

To show that \emph{s} is minimized with respect to the elements of
the weight vector, \emph{w}, given by Equation 4.2 it is necessary
to show that $\nicefrac{ds}{dW_{k,k}}$is zero for $k=1\,...\,q$.
To do this the expression $\left[I-AA^{T}W\right]TWA$ in Equation
4.14 will be considered. Applying Equations 4.10 and 4.11 gives:

\begin{equation}
\left[I-AA^{T}W\right]TWA=\left[I-AA^{T}W\right]A\varLambda=A\varLambda-A\varLambda=0
\end{equation}

This factor $\left[I-AA^{T}W\right]TWA$ of the expression on the
right hand side of Equation 4.14 is a matrix of zeroes which implies
that $\nicefrac{ds}{dW_{k,k}}$ is zero for all values of \emph{k}.
To show that this is a minimum and not a saddle point, it is necessary
to show that the corresponding Hessian matrix, \emph{H}, is positive
semi-definite.\newpage{} The elements of the Hessian matrix, $H_{i,j}$,
are given by: 

\begin{equation}
H_{i,j}=u^{T}AA^{T}\dfrac{dW}{dW_{i,i}}\left[I-AA^{T}W\right]T\dfrac{dW}{dW_{j,j}}AA^{T}u\quad\left(\mathrm{for}\:i,j=1\,...\,q\right)
\end{equation}

Using Equation 4.11 the matrix $\left[I-AA^{T}W\right]T$ can be expressed
as:

\begin{equation}
\left[I-AA^{T}W\right]T=T-A\varLambda A^{T}
\end{equation}

The $q\times q$ matrix $T-A\varLambda A^{T}$ is symmetric and it
has \emph{n} zero eigenvalues where \emph{n} is the number of columns
of the matrix \emph{A}. To show that the matrix $T-A\varLambda A^{T}$
is positive semi-definite it is sufficient to show that its smallest
nonzero eigenvalue, $\lambda_{\mathrm{min}}(q,n)$, is positive for
all values of \emph{q} and \emph{n}. For the case where $n=1$ the
smallest nonzero eigenvalue is given by:

\begin{equation}
\lambda_{\mathrm{min}}\left(q,1\right)=2\left[1-\cos\left(\dfrac{2\pi}{q+1}\right)\right]\quad\left(\mathrm{for}\:n=1,\:q=2,3,...\right)
\end{equation}

$\lambda_{\mathrm{min}}\left(q,1\right)$ is positive for all $q>1$.
If \emph{q} is fixed then the smallest nonzero eigenvalue increases
monotonically as \emph{n} increases from 1 to $q-1$. The value of
$\lambda_{\mathrm{min}}$ for $n=q-1$ is given by:

\begin{equation}
\lambda_{\mathrm{min}}\left(q,q-1\right)=4-\dfrac{2}{q+1}-\dfrac{2}{{2q \choose q}}\quad\left(\mathrm{for}\:n=q-1,\:q=2,3,...\right)
\end{equation}

where the parentheses on the right hand side denote a binomial coefficient.
Therefore the smallest nonzero eigenvalue is in the range $0<\lambda_{\mathrm{min}}(q,n)<4$
for all \emph{q} and $n<q$. This implies that the matrix $\left[I-AA^{T}W\right]T$
is positive semi-definite and can be factored as:

\begin{equation}
\left[I-AA^{T}W\right]T=R^{T}R
\end{equation}

The matrix \emph{S} is defined as:

\begin{equation}
S=R\,\mathrm{diag}\left(AA^{T}u\right)
\end{equation}

where the diag operator transforms the elements of a vector into the
elements of a diagonal matrix. With this the Hessian matrix, \emph{H,}
can be factored as:

\begin{equation}
H=S^{T}S
\end{equation}

Therefore \emph{H} is positive semi-definite and the weight vector
\emph{w}, given by Equation 4.2, minimizes the smoothing parameter
\emph{s.}

$ $

The matrix \emph{T} has the property that $-T$ is the operator that
acts on a vector to give the second differences of the vector. It
can be seen from Equation 4.2 that \emph{$-T$} acts on the vector
\emph{w} to yield $-v$, a vector of minus ones. This implies that
the elements of the vector \emph{w} can be generated by a quadratic
polynomial in the index \emph{i} whose leading term is $-\nicefrac{i^{2}}{2}$.
In fact the values of \emph{w} defined by Equation 4.2 are given by:

\begin{equation}
w_{i}=-\dfrac{i}{2}\left(i-q-1\right)\quad\left(\mathrm{for}\:i=1\,...\,q\right)
\end{equation}

This optimal weight function is a quadratic polynomial that is zero
at one sample interval beyond the \emph{q} sample intervals, i.e.:

\begin{equation}
w_{0}=w_{q+1}=0
\end{equation}
\newpage{}

\section{The Simplest Case of the S-G Filter}

To compare the relative performances of the standard S-G filter with
a constant weight function and the optimally weighted S-G filter with
a quadratic weight function the case of the simplest S-G filter will
be considered. In this case the fitting polynomial is just a constant.
The matrix \emph{X} of Equation 2.2 is a column vector of ones and
Equation 2.8 gives the filter coefficients:

\begin{equation}
c_{i}=\frac{1}{q}\quad\left(\mathrm{for}\:i=1\,...\,q\right)
\end{equation}

This is the definition of a moving average filter with \emph{q} samples.
The error reduction ratio (Equation 3.4) of the moving average filter
is given by:

\begin{equation}
r_{0}=\dfrac{\sigma_{\mathbf{e}}^{2}}{\sigma_{e}^{2}}=\frac{1}{q}
\end{equation}

where the subscript \emph{0} denotes a constant weight function. The
smoothing parameter (Equation 3.8) of a moving average filter is given
by:

\begin{align}
s_{0}=\dfrac{\sigma_{\Delta\mathbf{e}}^{2}}{\sigma_{\Delta e}^{2}}= & \frac{1}{q^{2}}
\end{align}

If Equation 2.10 is applied to a constant polynomial with the optimal
weight function given by Equation 4.23 then the filter coefficients
are:

\begin{equation}
c_{i}=\frac{6\,i\,(q+1-i)}{q(q+1)(q+2)}\quad\left(\mathrm{for}\:i=1\,...\,q\right)
\end{equation}

These coefficients are simply a constant multiple of the optimal weights.
The corresponding error reduction ratio is:

\begin{equation}
r_{2}=\frac{6}{5}\frac{(q+1)^{2}+1}{q\left(q+1\right)\left(q+2\right)}
\end{equation}

where the subscript \emph{2} denotes a quadratic weight function.
The corresponding smoothing parameter is given by:

\begin{equation}
s_{2}=\frac{6}{q\left(q+1\right)\left(q+2\right)}
\end{equation}

The constant residual weight function of the standard S-G filter has
the property of minimizing the error reduction ratio, \emph{r}. Therefore
it is not possible to select residual weights which will minimize
both the error reduction ratio, \emph{r}, and the smoothness parameter,
\emph{s}. The following two approximations illustrate the nature of
the tradeoff for the case where the S-G fitting polynomial is a constant:

\begin{equation}
\dfrac{r_{0}}{r_{2}}\approx\dfrac{5}{6}\left(1+\dfrac{1}{q}\right)\quad\left(\mathrm{for}\:q>5\right)
\end{equation}

\begin{equation}
\dfrac{s_{0}}{s_{2}}\approx\dfrac{q}{6}\left(1+\dfrac{3}{q}\right)\quad\left(\mathrm{for}\:q>3\right)
\end{equation}

The error of these approximations approaches zero with increasing
values of \emph{q}. So as the sample size, \emph{q}, increases the
relative advantage of a constant weight function minimizing the error
reduction ratio,\emph{ r}, approaches a constant whereas the relative
advantage of a quadratic weight function minimizing the smoothness
parameter, \emph{s}, approaches proportionality to \emph{q}. This
asymmetry in the tradeoffs heavily favours optimizing the smoothing
parameter, \emph{s}, with a quadratic weight function rather than
optimizing the error reduction ratio, \emph{r}, with a constant weight
function.

\section{The General Case of the S-G Filter}

For the general case of the S-G filter, fitting polynomials of degree
higher than zero will be considered. However the analysis here will
be limited to the case where the middle element of the \emph{u} vector
of Equation 2.6 is chosen to be one, i.e.,$u_{\nicefrac{(1+q)}{2}}=1$.
It is this case for which the coefficient vector \emph{c} is symmetric
and the S-G filter is linear phase.

This case also has the counterintuitive property that only even polynomials
need be considered in the derivation of the S-G filter coefficients.
In this case the degree of the fitting polynomial is even. As before,
\emph{n} is the number of columns of the matrix \emph{X} in Equation
2.2 where the \emph{j}th column is $x^{2j-2}$. It will also be convenient
to denote the index of the middle element of the vector \emph{u} by
\emph{m} so $m=\nicefrac{(1+q)}{2}$ and $u_{m}=1$.

Section 5 dealt with the simplest case of the S-G filter where the
fitting polynomial is a constant. Even for this simplest case, Equations
5.5 and 5.6 for\emph{ $r_{2}$} and \emph{$s_{2}$} are not simple.
And these equations increase in complexity with increasing degree
of the fitting polynomial. For the purpose of comparing the relative
performance of the S-G filter with a constant and a quadratic weight
function good approximations to the ratios $\nicefrac{r_{0}}{r_{2}}$
and $\nicefrac{s_{0}}{s_{2}}$ are adequate to illustrate how these
ratios are influenced by the values of the filter parameters \emph{m}
and \emph{n:}

\begin{equation}
\dfrac{r_{0}}{r_{2}}\approx1-\dfrac{\left(1-\frac{n}{m}\right)^{2}}{2\left(2n+1\right)}
\end{equation}

\begin{equation}
\dfrac{s_{0}}{s_{2}}\approx1+\dfrac{3m\left(1-\frac{n}{m}\right)^{2}}{\left(2n+1\right)^{2}}
\end{equation}

For the case where $m=n$ these approximations give the exact result
that $\nicefrac{r_{0}}{r_{2}}=\nicefrac{s_{0}}{s_{2}}=1$. In fact,
in this case the S-G equations yield a degenerate solution where there
is only one nonzero filter coefficient and the filter output is equal
to the filter input. For a practical filter this imposes the constraint
that $m>n$. The conclusion drawn in Section 5 for the simplest case
of the S-G filter where the fitting polynomial is a constant also
holds for the general case of polynomials of any degree. The modest
advantage of a constant weight function in the error reduction ratio,
\emph{r}, is generally outweighed by a significant advantage of a
quadratic weight function in the smoothness parameter, \emph{s}.

This result can also be interpreted in the frequency domain. The S-G
filter cutoff frequency is very nearly proportional to the error reduction
ratio, \emph{r}. Therefore the cutoff frequency is relatively insensitive
to the choice of residual weight function while the frequency response
in the stopband is much more sensitive to the choice of residual weight
function {[}3{]}.

$ $

For the sake of completeness, Turton's use of a triangular residual
weight function {[}3{]} will be considered. The triangular weight
function, \emph{wt}, is given by: 

\begin{equation}
wt_{i}=1-\left|1-\dfrac{2i}{q+1}\right|\quad\left(\mathrm{for}\:i=1\,...\,q\right)
\end{equation}

If the smoothness parameter corresponding to this triangular weight
function is denoted by $s_{1}$ then the filter smoothness parameter
ratio, $\nicefrac{s_{0}}{s_{1}}$, is approximated by:

\begin{equation}
\dfrac{s_{0}}{s_{1}}\approx1+\dfrac{3m\left(1-\frac{n}{m}\right)^{2}}{\left(2n+\nicefrac{3}{2}\right)^{2}}
\end{equation}

A comparison with Equation 6.2 giving the smoothness parameter ratio,
$\nicefrac{s_{0}}{s_{2}}$, for a quadratic weight function shows
that the triangular weight function is very close to being optimal.\newpage{}

\section{Conclusion}

In his 1992 paper {[}3{]} Turton showed that applying a triangular
residual weighting function in the derivation of the S-G filter significantly
improves the frequency response of the S-G filter in the stopband.
However Turton's analysis has been largely overlooked in expository
discussions of the S-G filter {[}2{]}, {[}4{]}, {[}5{]}. This paper
has shown that a quadratic residual weight function optimizes a particular
measure of filter output smoothness. While this may be an interesting
theoretical result, it represents only a slight improvement over Turton's
choice of a triangular residual weight function.

The practical contribution of the present paper is in emphasizing
the significance of Turton's result with a time-domain analysis of
the smoothing property of the S-G filter. Turton concluded his paper
with this recommendation for the use of his variant of the S-G filter:
``It should therefore be used in preference to the Savitzky-Golay
filter in future spectroscopic applications.'' The analysis presented
here is offered in support of Turton's recommendation.

\section*{References}

{[}1{]} Savitzky, A. and Golay M.J.E. ``Smoothing and differentiation
of data by simplified least-squares

\qquad{}procedures'' \emph{Anal. Chem.} 36 (1964) 1627-39

{[}2{]} Schafer, R.W. ``What is a Savitzky-Golay filter?'' \emph{IEEE
Signal Process. Mag}. July 2011, 111-117

{[}3{]} Turton, B.C.H. ``A novel variant of the Savitzky-Golay filter
for spectroscopic applications''. 

\qquad{}\emph{Meas. Sci. Technol.} $\mathbf{3}$ (1992) 858-86

{[}4{]} Orfanidis, S.J. ``Introduction to Signal Processing'' Prentice
Hall, 1995

{[}5{]} Press, W.H et. al. ``Numerical Recipes, 3rd Ed.'' Cambridge
University Press, 2007
\end{document}